\documentclass[aps,prb,reprint,superscriptaddress]{revtex4-2}
\usepackage{graphicx}
\usepackage{dcolumn}
\usepackage{bm}
\usepackage[utf8]{inputenc}
\usepackage[T1]{fontenc}
\usepackage[T1]{fontenc}
\usepackage{newtxtext,newtxmath}
\usepackage{textcomp}
\usepackage{mathrsfs}
\usepackage{gensymb}
\usepackage{textgreek}
\usepackage{amsmath}
\usepackage{natbib}
\usepackage{color}
\usepackage{hyperref}
\usepackage{bm}
\usepackage[caption=false]{subfig}
\usepackage{verbatim}
\usepackage{ulem}
\usepackage{soul}
\usepackage{array,makecell,tabularx,booktabs}

\DeclareUnicodeCharacter{0308}{\"}

\begin{document}

\title{Nitrogen-Vacancy-Mediated Magnetism in Sputtered GdN Thin Films}
\author{Pankaj Bhardwaj}
\email{pankajb@iisc.ac.in}
\affiliation{Centre for Nanoscience and Engineering, Indian Institute of Science, Bengaluru 560012, India} 
\author{Jyotirmoy Sarkar}
\affiliation{Theoretical Sciences Unit, Jawaharlal Nehru Center for Advanced Scientific Research, Jakkur, Bengaluru 560064, India}
\author{Bubun Biswal}
\affiliation{Department of Physics, Indian Institute of Technology Madras, Chennai 600036, India}
\author{Subhransu Kumar Negi}
\affiliation{Centre for Nanoscience and Engineering, Indian Institute of Science, Bengaluru 560012, India}
\author{Arijit Sinha}
\affiliation{Theoretical Sciences Unit, Jawaharlal Nehru Center for Advanced Scientific Research, Jakkur, Bengaluru 560064, India}
\author{Anirudh Venugopalrao }
\affiliation{Centre for Nanoscience and Engineering, Indian Institute of Science, Bengaluru 560012, India}
\author{Sharath Kumar C}
\affiliation{School of Physics, IISER Thiruvananthapuram, Vithura, Thiruvananthapuram 695551, India}
\author{Sreelakshmi M Nair}
\affiliation{Department of Physics, BITS Pilani K. K. Birla Goa Campus, Zuarinagar 403726, Goa, India}
\author{R. S. Patel}
\affiliation{Department of Physics, BITS Pilani K. K. Birla Goa Campus, Zuarinagar 403726, Goa, India}
\author{Deepshika Jaiswal Nagar}
\affiliation{School of Physics, IISER Thiruvananthapuram, Vithura, Thiruvananthapuram 695551, India}
\author{Abhishek Mishra}
\affiliation{Department of Physics, Indian Institute of Technology Madras, Chennai 600036, India}
\author{Srinivasan Raghavan}
\affiliation{Centre for Nanoscience and Engineering, Indian Institute of Science, Bengaluru 560012, India}
\author{Umesh Waghmare}
\affiliation{Theoretical Sciences Unit, Jawaharlal Nehru Center for Advanced Scientific Research, Jakkur, Bengaluru 560064, India}
\author{Dhavala Suri}
\email{dsuri@iisc.ac.in}
\affiliation{Centre for Nanoscience and Engineering, Indian Institute of Science, Bengaluru 560012, India}

\begin{abstract}
Among rare-earth nitrides (RENs), gadolinium nitride (GdN) stands out as a promising material for spintronics owing to its distinctive combination of semiconducting behavior, strong exchange interactions, and intrinsically soft ferromagnetism. Its relatively high Curie temperature and large saturation magnetization make it an attractive candidate for device concepts such as non-volatile memory elements and spin-based transistors, motivating efforts toward low-cost, uniform, and compositionally controlled thin-film growth. In this work, we deposited GdN thin films on SiO$_2$/AlN substrates using DC sputtering under reactive nitridation conditions, with thicknesses varying from 18 to 180 nm, and systematically investigated their structural and magnetic properties. The films exhibit soft ferromagnetic ordering, characterized by a coercive field of approximately 200 Oe and a Curie temperature ($T_{\mathrm{c}}$) near 70 K. Structural analysis reveals lattice distortions and local strain associated with nitrogen-vacancy defects, whose concentration varies with film thickness. Our theoretical studies establish a direct correlation between the observed Raman modes of the GdN lattice and the reduced magnetization induced by nitrogen vacancies. These vacancies give rise to defect-mediated ferromagnetism, leading to a measurable enhancement of $T_{\mathrm{c}}$ from 68 K to 82 K across the studied thickness range. The observed magnetic behavior is well described by the bound magnetic polaron (BMP) model, confirming that nitrogen vacancies are key contributors to ferromagnetic ordering while preserving the soft-magnetic character intrinsic to GdN. This study underscores the pivotal role of defect engineering in optimizing GdN thin films for spintronics applications.
\end{abstract}

\maketitle

\section{Introduction}
    Rare-earth nitrides (RENs) have recently garnered significant attention due to their diverse applications in various technologies, particularly in the fields of magnetism and electronics. Theoretical and experimental investigations have elucidated that these properties arise from the complex interaction of partially filled 4\textit{f }electrons.\cite{Holmes-Hewett2025-sw} The 4\textit{f} shell elements possess higher angular quantum numbers compared to \textit{p} and \textit{d} electron shells, enabling RENs to have both orbital and spin degrees of freedom, thus they exhibit robust magnetic moments and ferromagnetic (FM) feature. Most rare earth (RE) elements exist in a stable +3 charge state and readily react with nitrogen to form RENs, which consist of rare-earth (RE$^{3+}$) and nitrogen (N$^{3-}$) atoms arranged in ABAB...pattern, exhibiting NaCl-type cubic rock salt structure.\cite{Kneisel2024-yu} During the bonding process with the N atom, the RE atom loses its 5\textit{d} electron, which leaves only strong localization of \textit{f}-orbitals in the valence band and results in intrinsic ferromagnetism.\cite{Larson2007-jn} RENs materials exhibit semiconducting and magnetic properties simultaneously, making them ferromagnetic semiconductors (FMSc). These materials can harness both charge carriers and electron spins, making them suitable for both applied and fundamental research, making them promising candidates for next-generation spintronics devices.\cite{Dietl2014-wz} The limitations of material growth, particularly the high susceptibility of nitride materials to the ambient environment and their greater affinity for oxygen, pose significant challenges to the advancement of research on RENs materials.\cite{Natali2013-xs} With the advancement of ultra-high vacuum (UHV) deposition systems in the early 1980s, including molecular beam epitaxy (MBE), pulsed laser deposition (PLD), and UHV sputtering, the research progress of REN materials was significantly accelerated. RENs materials are extensively utilized in numerous everyday consumer applications, ranging from displays,\cite{McKay2020-rs} lighting systems,\cite{Xie2010-jx} hydrogen liquefaction,\cite{Tian2025-qa} electronics,\cite{Anton2016-wl}and quantum devices.\cite{Chang2013-sr}\cite{Spellman2022-cm}

 Among various RENs, gadolinium nitride (GdN) has garnered substantial attention for potential spintronics device fabrication due to its intrinsic spin magnetic moment.\cite{Leuenberger2005-iy} The bulk GdN crystallizes as a NaCl-type cubic structure, with a lattice constant of 4.98 Å,\cite{Kneisel2024-yu} where all gadolinium (Gd) ions possess a +3 charge state having a ground state of 8\textit{S}$_{7/2}$ and 4\textit{f}$^7$ electronic configuration. This ground-state interatomic 4\textit{f}-5\textit{d} exchange interaction contributes solely to spin moment, resulting in a substantial net magnetic moment of 7.0 μ$_b$/Gd$^{3+}$ ions.\cite{Larson2007-jn} GdN exhibits the highest Curie temperature (\textit{T$_{c}$)} of 72 K among RENs materials, making it an attractive candidate for the development of high-temperature spintronics devices.\cite{Plank2011-po} Furthermore, GdN exhibits an optical bandgap of 1.7 eV in the ferromagnetic regime and exhibits half-metallic behavior, low coercivity, and strong spin polarization, which makes GdN a perfect candidate as a spin-filter.\cite{Vilela2024-mn}  This process finds applications in spin valves, magnetic tunnel junctions (MTJs), spin transistors, and spin-based quantum computing. Despite the significant potential of GdN thin films, the fabrication and growth of high-quality GdN are challenging due to inherent defects, such as nitrogen vacancies and high oxophilicity. GdN easily reacts with atmospheric oxygen, forming Gd$_2$O$_3$, which substantially impacts its electronic and magnetic properties.\cite{Ludbrook2009-oh} 
 
 GdN thin films exhibit a strong sensitivity of their structural and magnetic responses to growth conditions, particularly stoichiometry and the concentration of nitrogen vacancies. Early reactive magnetron sputtering studies conducted at base pressures of $2 \times 10^{-9}$ mbar identified two distinct structural–magnetic regimes, commonly referred to as GdN-I and GdN-II. The GdN-I phase displays a Curie temperature ($T_c$) of approximately 60~K, consistent with near-stoichiometric Gd:N ratios (1:1) and the expected ferromagnetic order of bulk-like GdN. In contrast, the GdN-II phase emerges under nitrogen-deficient conditions and is associated with pronounced nitrogen vacancies ($V_{\mathrm{N}}$) formation, leading to deviations from ideal stoichiometry. This phase has been linked to modified electronic density and lattice distortions, effects that were originally interpreted as giving rise to antiferromagnetic tendencies \cite{Senapati2011-bq}. Complementary synthesis efforts using chemical vapor deposition with AlN buffer layers produced polycrystalline GdN films exhibiting a reduced $T_c$ of $\sim 30$~K, an effect attributed to strain-induced lattice expansion \cite{Cwik2017-vf}. More recent molecular beam epitaxy studies, in which the nitrogen pressure was systematically varied, revealed the coexistence of GdN-I and GdN-II phases similar to the radio-frequency magnetron sputtering (RFMS)-grown films; however, these reports demonstrate that increasing $V_{\mathrm{N}}$ enhances $T_c$ and strengthens both magnetization and coercivity, challenging earlier assertions of intrinsic antiferromagnetism in the GdN-II phase \cite{Shaib2020-ox}. Sputtered films grown under ultrahigh vacuum conditions ($5 \times 10^{-8}$~torr) further highlight the role of lattice expansion, exhibiting a $T_c$ of $\sim 30$~K and a reduced moment of 4.5~$\mu_B$/Gd$^{3+}$, consistent with suppression of the magnetic moment due to structural dilation \cite{Vilela2024-mn}.

In this work, we investigate the growth and optimization of GdN thin films on SiO$_2$/AlN substrates using DC sputtering at a base pressure of $10^{-7}$\,mbar. The deposition protocol incorporates substrate degassing and controlled pre- and post-nitridation steps to promote stoichiometric GdN formation. Structure by XRD, Raman spectroscopy, XPS, and TEM was performed across a range of film thicknesses to examine the evolution of nitrogen vacancies and the s and the resulting emergence of the two characteristic GdN phases. Magnetic measurements as a function of temperature and field further elucidate the influence of nitrogen deficiency on the magnetization and Curie temperature of the films.

\begin{figure*}[!ht]
    \centering
    \includegraphics[width=\textwidth]{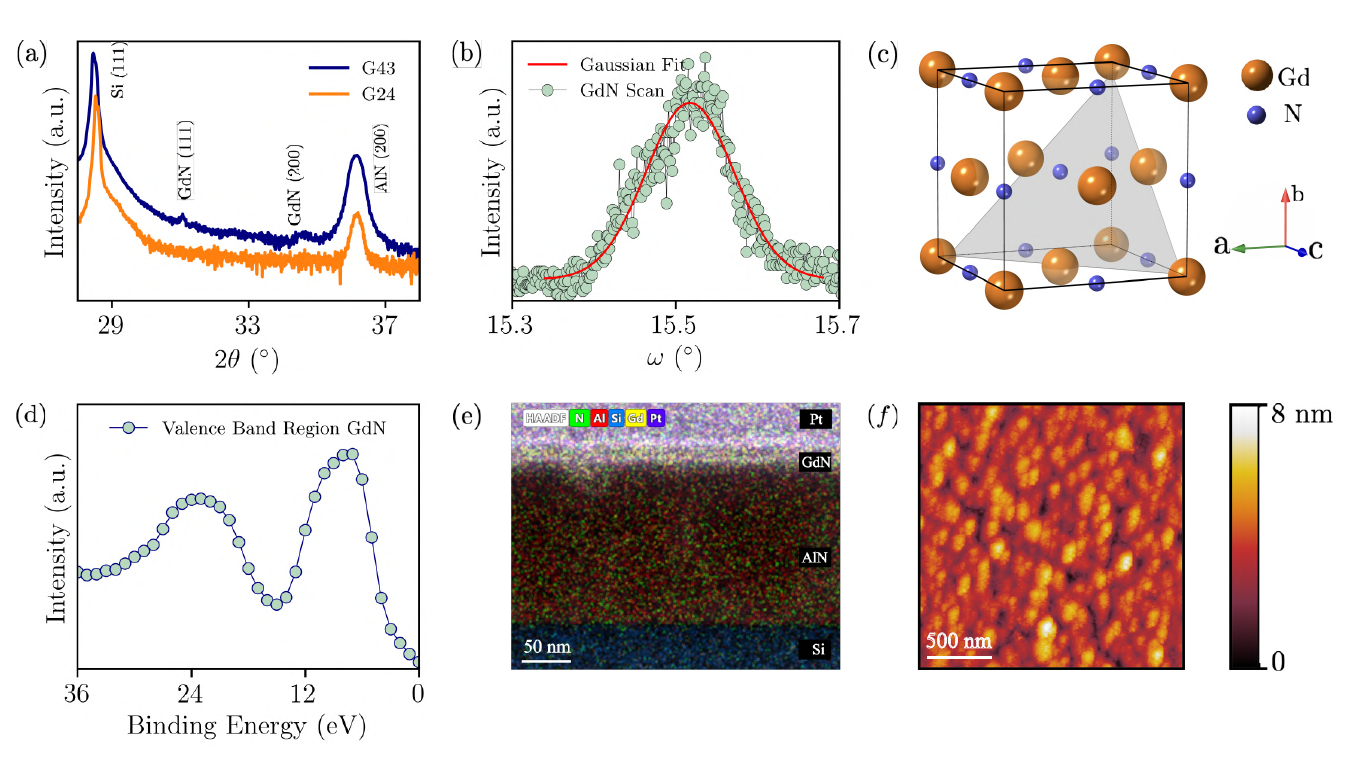} % Replace with your filename
    \caption{ (a) The 2 theta ($\theta)$- omega ($\omega)$ scan of the X-ray diffraction (XRD) pattern of the G24 and G43 sample, where G24 shows non-growth and G43 shows growth of GdN thin of (111) plane film deposited on an AlN/SiO$_2$ substrate (indicates the effect of nitridation during thin film growth). (b) The $\omega$-scan of the GdN (111) reflection, fitted with a Gaussian function having a full width at half maximum (FWHM) of 0.12, confirmed the crystalline, textured thin films with an incursion of dislocation density of 2.84 $\times$ 10\textsuperscript{9} cm\textsuperscript{2}. (c) The cubic crystal lattice of the grown GdN thin film, analyzed using Crystal Diffract software and modeled by Crystal Maker software, exhibits a NaCl-type cubic structure, where Gd and N atoms are arranged in an ABAB... pattern, having (111) reflection, with a lattice spacing of 3.529 \AA{}.  (d) The X-ray photoelectron (XPS) depicts the valence band region, revealing the energy positions of the Gd 5\textit{p} orbitals at 21 eV and 4\textit{f} orbitals at 7.5 eV correspond to the multiplet final state of 4\textit{$f^6$} Gd$^{3+}$ ions, which are situated below the Fermi level, thereby confirming the successful growth of GdN thin films. (e) Cross-sectional transmission electron microscopy energy dispersive spectroscopy (TEM-EDS) mapping images shows the GdN growth in AlN/SiO$_2$ substrate (f) Atomic force microscopy (AFM) image of the deposited GdN thin film of 2 $\times$ 2 \textmu{}m\textsuperscript{2} confirms smooth and homogeneous growth, with a roughness (R$_q$) of approximately 0.9 nm.}
    \label{intro}
\end{figure*}

 \section{Synthesis}

  GdN thin films of varying thicknesses were deposited on 3.5~$\times$~3.5~mm$^2$ Si/SiO$_2$ (280~nm) substrates using confocal DC magnetron sputtering. In contrast to earlier reports, which predominantly relied on base pressures of $10^{-9}$--$10^{-10}$~mbar, the present work develops an optimized and cost-effective growth protocol at a base pressure of $2.5 \times 10^{-7}$~mbar, enabling scalable GdN deposition for spintronic device applications. To suppress interfacial reactions between GdN and Si, a 180~nm AlN buffer layer was first deposited by CVD, preventing the formation of GdSi$_2$ and subsequent substrate degradation \cite{Natali2012-xc}. The cleaned AlN/SiO$_2$/Si substrates were mounted in the sputtering chamber, degassed at 400\,$^\circ$C, and pre-sputtered in high-purity Ar (5N). As Gd acts as an efficient getter, this step reduces residual gas contamination and stabilizes the vacuum environment. Film deposition was carried out at a working pressure of $8 \times 10^{-4}$~mbar using a controlled Ar/N$_2$ mixture, with a target–substrate distance of 7.8~cm and substrate rotation at 5~rpm to ensure uniform thickness and minimize plasma heating. The substrate temperature was maintained at 500\,$^\circ$C, and a DC power of 100~W was applied during GdN growth.

Despite following the initial deposition protocol, the formation of GdN was insufficient, as evidenced by the G24 sample (orange curve) in Fig.~1~(a). This limitation stems from the large mass mismatch between Gd and N species in the sputtering plasma, which necessitates a substantially higher N$_2$ flux—often two or more orders of magnitude greater—to achieve stoichiometric GdN \cite{Ullstad2019-zh}. Consequently, precise tuning of the N$_2$ concentration was required (see SI for growth parameters). To facilitate GdN formation, a post-deposition nitridation step was employed, wherein the films were annealed at 500\,$^\circ$C under an N$_2$ partial pressure of $9 \times 10^{-3}$\,mbar within the same UHV system \cite{Yadav2021-dn}. During nitridation, nitrogen decomposition and diffusion are promoted by the strong Gd–N affinity, leading predominantly to the formation of stoichiometric cubic GdN (GdN-I), accompanied by a minority GdN-II phase. As these processes are strongly temperature dependent, nitridation kinetics play a key role in phase stabilization. For thickness-optimization studies, additional GdN films were grown under identical conditions with deposition times varied between 15 and 240\,s.

\section{Structural Characterization} The structural characteristics of the sputtered GdN thin films were investigated by X-ray diffraction (XRD). Fig.~1~(a) displays the $2\theta$--$\omega$ scans of the reference nongrown sample (G24) and the optimized GdN-grown sample (G43). The diffraction features at approximately $28.5^\circ$ and $27.1^\circ$ originate from the Si(111) substrate and the native SiO$_2$ layer, respectively, while the pronounced peak at $36 \pm 0.1^\circ$ corresponds to the AlN(100) buffer layer with an estimated thickness of $\sim160$\,nm. A distinct reflection observed at $31 \pm 0.1^\circ$ confirms the formation of cubic GdN with a preferential (111) orientation. In addition, the peak near $34.5^\circ$ is assigned to the GdN(200) plane, in good agreement with previous reports~\cite{Anton2023-gi}. The simultaneous presence of the (111) and (200) reflections indicates polycrystalline GdN growth on the AlN/SiO$_2$ substrate, consistent with the standard JCPDS card 00-015-0888 for cubic GdN (space group Fm$\bar{3}$m).

Fig.~1~(b) shows the $\omega$-scan of the GdN(111) reflection centered at $15.55^\circ$. A Gaussian fit to the rocking curve yields a full width at half maximum (FWHM) of $0.127^\circ$ (see Fig.~S1 in the Supplementary Information), indicating good crystalline quality with a moderate mosaic spread. The corresponding threading dislocation density ($\rho$) was estimated following Ref.~\cite{Yan2024-qj} using
\begin{equation}
\rho \approx \frac{\beta^{2}}{4.35\,b^{2}},
\end{equation}
where $\beta$ is the rocking-curve FWHM expressed in radians and $b$ is the magnitude of the Burgers vector. Using this relation, a dislocation density of $\rho \approx 2.84 \times 10^{9}$\,cm$^{-2}$ was obtained. Fig.~1~(c) illustrates the cubic crystal structure of GdN, analyzed using \textit{Crystal Diffract} software based on the experimentally observed XRD pattern and modeled using \textit{CrystalMaker}. The cubic lattice clearly resolves the spatial separation of the Gd$^{3+}$ and N$^{3-}$ ions, with an interatomic bond length of 2.496~\AA\ and an interplanar spacing of 3.529~\AA\ corresponding to the (111) planes. The stacking sequence follows an ABABAB\,$\ldots$ arrangement, consistent with the pronounced electronegativity difference between Gd and N ions, resulting in a strong ionic and nearly isotropic bonding character along the (111) direction~\cite{Ludbrook2009-oh}. The extracted lattice parameter for the GdN thin film is 4.990~\AA, which is in close agreement with the reported bulk lattice constant of 4.98~\AA~\cite{Li2015-dk}. To further quantify the structural characteristics, the lattice misfit strain ($M\%$) between the GdN thin film and the AlN buffer layer was evaluated using
\begin{equation}
\text{Lattice misfit (M\%)} = \frac{a_{\text{film}} - a_{\text{sub}}}{a_{\text{sub}}} \times 100,
\end{equation}
where $a_{\text{film}}$ and $a_{\text{sub}}$ denote the lattice constants of the GdN thin film and the AlN buffer layer, respectively. The calculated lattice misfit is $13.54\%$, indicating a substantial lattice mismatch that is predominantly accommodated through strain relaxation mechanisms and contributes to the observed polycrystalline growth of GdN on AlN.

\begin{figure*}[!ht]
\centering
\includegraphics[width=\textwidth]{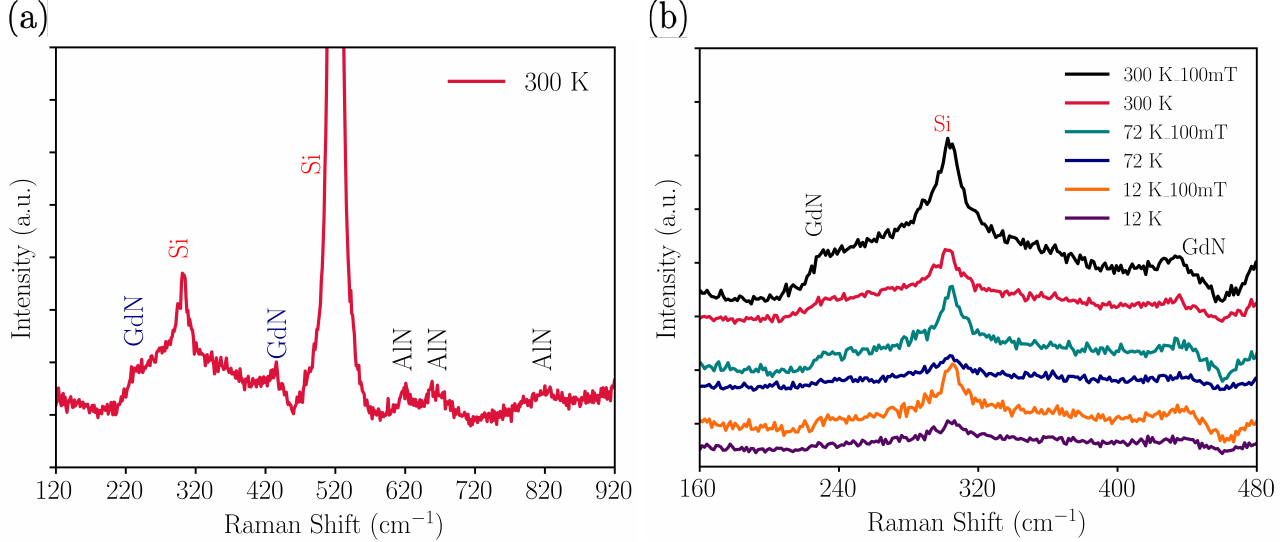} 
\caption{(a) Raman spectra of a grown cubic GdN thin film at room temperature, exhibiting two distinct frequency modes at 235 and 435 cm$^{-1}$, corresponding to the TO (Γ) and LO (X) modes of GdN, respectively. (b) In the absence of a magnetic field, it is observed that the decrease in temperature causes the frequency mode of the GdN shoulder to broaden, resulting in the disappearance of the Raman mode at 12 K, attributed to lattice strain in the grown thin films. Conversely, with an applied magnetic field of 100 mT, the cubic symmetry of GdN is further broken, leading to an increased shoulder; however, it follows the same pattern of broadening with a reduction in temperature.}
\label{raman}
\end{figure*}

However, the intrinsic lattice mismatch between AlN and GdN is approximately $\sim13\%$. Deviations from the ideal mismatch are likely influenced by high-pressure sputtering conditions and the specific growth parameters employed, which can introduce an additional $\sim2\%$ excess misfit strain in the GdN thin film. This excess strain contributes to enhanced lattice distortion and increased dislocation density, thereby promoting the formation of nitrogen vacancies (\textit{V$_\mathrm{N}$})~\cite{Holmes-Hewett2020-kl}. Despite optimized growth conditions, the deposited GdN films do not exhibit complete (111) texturing. A detailed analysis of the XRD profiles reveals a gradual peak broadening accompanied by asymmetric line shapes, indicating the presence of structural disorder. Notably, the GdN (200) reflection exhibits a slight shift toward $\sim34.5^\circ$, which is indicative of cubic lattice distortion. The observed asymmetry in the diffraction peaks is attributed to the formation of a condensed GdN phase, commonly referred to as GdN-II. The emergence of the GdN-II phase is associated with the large mass disparity between Gd and N ions, leading to significant differences in their kinetic energies within the sputtering plasma~\cite{Kneisel2025-ao}. This imbalance results in a sensitive dependence of film stoichiometry on the relative Gd and N fluxes, giving rise to a structurally distorted, nitrogen-deficient GdN phase (GdN-II) that retains a lattice parameter close to that of stoichiometric GdN.

Fig.~1~(e) shows transmission electron microscopy energy-dispersive spectroscopy (EDS) mapping images of the grown GdN thin film, illustrating a uniform GdN layer situated between the Pt and the buffer AlN layer. The surface morphology of the GdN thin films was examined by atomic force microscopy (AFM) operated in tapping mode, as shown in Fig.~1~(f). The scan over a $2 \times 2~\mu\text{m}^2$ area reveals a smooth, homogeneous, and uniformly grown film with a root-mean-square roughness ($R_q$) of approximately 0.9~nm, indicative of high surface quality. To further probe the chemical composition and bonding environment, high-resolution X-ray photoelectron spectroscopy (XPS) measurements were performed on the GdN films deposited on AlN/SiO$_2$ substrates. Owing to the surface-sensitive nature of XPS, depth profiling was carried out using Ar$^+$ ion etching to a depth of $\sim5$~nm to remove the native surface layer and access the underlying film. The wide-energy survey spectrum (0--1300~eV) confirms the presence of Gd, Al, and N species in the deposited films, as shown in Fig.~S2~ of SI.  Fig.~1~(e) presents the valence band spectrum, dominated by contributions from the Gd $4f$ and $5p$ states. While theoretical studies predict the Gd $4f$ ground-state density of states to lie below the Fermi level in stoichiometric GdN~\cite{Larson2006-jr}, the experimentally observed broadened multiplet structure corresponding to the Gd$^{3+}$ ($4f^6$) final state appears at a binding energy of $7.5 \pm 0.5$~eV. This feature is shifted by approximately 0.8~eV toward higher binding energy, in good agreement with previously reported results and providing further confirmation of GdN formation~\cite{Leuenberger2005-iy,Kneisel2025-ao}.

\section{Raman Analysis}
To probe the structural evolution of the GdN thin films grown on AlN, temperature-dependent Raman spectroscopy was performed over the range of 10--300~K, both in the presence and absence of an external magnetic field,as shown in Fig.~S3~ of SI. Rare-earth nitrides (RENs) crystallize in the centrosymmetric rocksalt (NaCl-type) structure, for which first-order Raman scattering is symmetry forbidden due to inversion symmetry at both the cation and anion sites~\cite{Shaib2020-ox}. Nevertheless, several RENs have been reported to exhibit Raman-active features arising from local symmetry breaking induced by nitrogen vacancies, lattice disorder, or strain~\cite{Granville2009-au}.
Fig.~2~(a) shows the room-temperature Raman spectrum of the grown cubic GdN thin film. The measured spectrum is dominated by contributions from the Si, SiO$_2$, and AlN substrate layers. The intrinsic longitudinal optical (LO) phonon modes of GdN are expected to appear in the range of $\sim$500--550~cm$^{-1}$. However, in the present case, the strong Si Raman mode at 521~cm$^{-1}$, accompanied by a relatively broad linewidth, masks the weaker GdN-related signal in this spectral window. Notably, two broad shoulder-like features observed at approximately 235~cm$^{-1}$ and 435~cm$^{-1}$ can be assigned to infrared-active transverse optical (TO, $\Gamma$) and longitudinal optical (LO, X) phonon modes of GdN, respectively. These features are consistent with previously reported experimental observations and theoretical predictions, while the remaining Raman peaks originate from the underlying Si and AlN layers~\cite{Van_Koughnet2023-np,Holmes-Hewett2020-kl}.

Fig. 2~(b) presents the temperature evolution of the Raman spectra recorded between 300 and 10 K under an applied magnetic field of 100 mT. Upon cooling, the shoulder associated with the GdN longitudinal optical (LO) phonon mode exhibits a progressive broadening, observed both in the absence and presence of the magnetic field. A pronounced enhancement of this broadening occurs near 72 K, corresponding closely to the Curie temperature ($T_\mathrm{c}$) of GdN. The gradual suppression of the LO-related feature with decreasing temperature is attributed to the reduction of lattice fluctuations, which diminishes the nitrogen-vacancy–activated Raman response and eventually leads to its disappearance~\cite{Van_Koughnet2023-np}. Notably, under an applied magnetic field, the suppressed LO mode becomes more pronounced compared to the zero-field case. This magnetic-field-induced enhancement is attributed to the coupling between lattice disorder, nitrogen vacancies, and the formation of magnetic polarons in both the paramagnetic and ferromagnetic regimes, which further lifts the cubic symmetry constraints of GdN~\cite{Eiter2014-ae,Shao2016-hq}. Such coupling enhances electron–phonon scattering, giving rise to a broadened and diffusive LO-mode shoulder. Near $T_\mathrm{c}$ ($\sim$70 K), the onset of spin alignment intensifies this effect, leading to substantial broadening and partial flattening of the Raman feature. At lower temperatures (10 K), the spins are largely aligned and thermal fluctuations are suppressed, resulting in a flattened LO response. Nevertheless, a weak residual shoulder persists due to the interaction of remaining paramagnetic contributions with the applied magnetic field~\cite{Ogita2003-hv}.

\begin{figure}[!ht]
\centering
\includegraphics[width=1\linewidth]{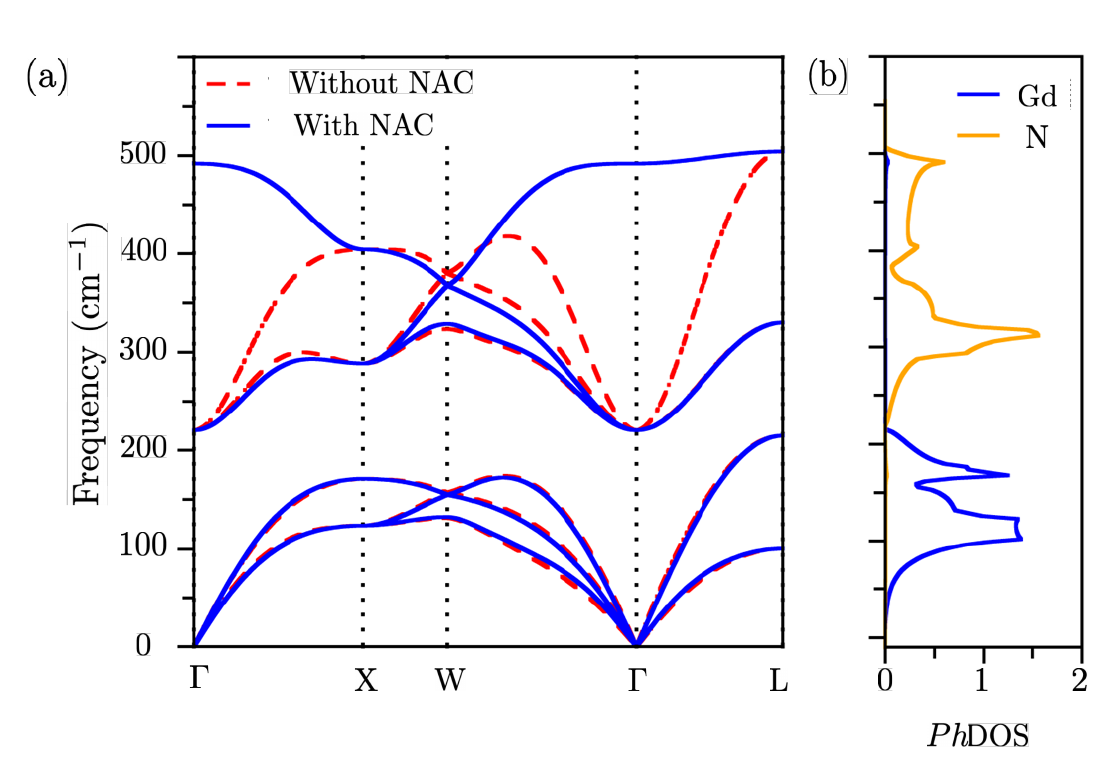} 
\caption{(a) The computed phonon dispersion of GdN lattice within Brillouin zone (b) The Gd and N atom projected vibrational density of states per cm$^{-1}$.}

\label{Raman DFT}
\end{figure}

The foremost puzzle here is that the cubic Rocksalt structure of any compound can NOT have Raman active modes. However, defects can give rise to nontrivial peaks in the Raman spectra that are seen in experiment. In GdN, we find that N-vacancy can give rise to Raman peaks, and identify the specific modes that are responsible for them: specific phonons
at X and R points. First-principles density functional theory (DFT) calculations were conducted utilizing the Vienna Ab initio Simulation Package (VASP) code\cite{Kresse1996_1,Kresse1996}. Phonon dispersion was determined employing the frozen phonon method with a 2 × 2 × 2 supercell for bulk GdN, while density functional perturbation theory (DFPT) was used to calculate Γ-point phonon frequencies of 1 x 1 x 1 cell of defected GdN. Experimental and relaxed lattice parameters of the conventional unit cell of GdN are 4.99 and 5.09 \AA, respectively (additional information regarding electronic and vibrational properties is provided in the SI).
Phonon spectrum exhibits well-defined acoustic and optical branches with no imaginary frequencies across the Brillouin zone, therey confirming the dynamical stability of the rocksalt GdN structure. Long-range elcetrostatic interactions cause a distinct longitudinal optical-transverse optical (LO-TO) splitting near Γ-point. Frequencies of zone-center optical modes are in good agreement with previously reported theoretical data \cite{Van_Koughnet2023-np}.

A supercell of defective GdN containing 63 atoms has 189 phonon modes at the Γ-point. Notably, two optical phonons at 240~cm$^{-1}$ and 242~cm$^{-1}$ comparable to the experimentally observed Raman peak at 235~cm$^{-1}$, correspond to $T_{2g}$ and $A_{1g}$ irreducible representations, respectively. Additionally, a phonon mode at 428~cm$^{-1}$, close to the experimentally observed Raman peak at 435~cm$^{-1}$, belongs to $E_{g}$ irreducible representation (visualization of modes added in SI). N-vacancy in the GdN crystal structure allows for absorption of crystal momentum, relaxing the momentum conservation rule (requiring nearly zero phonon momentum) governing Raman spectra \cite{Venezuela2011}. This facilitates the participation of phonons at q $\neq$ 0 in the Raman scattering process, thereby accounting for the experimentally observed Raman modes.%Visualization of A1g and Eg modes of GdN1-x (x= 0.03125) shows that they are composed of the modes at L and X points (   ) of the BZ in the figure. 

\begin{figure}[!ht]
\centering
\includegraphics[width=1\linewidth]{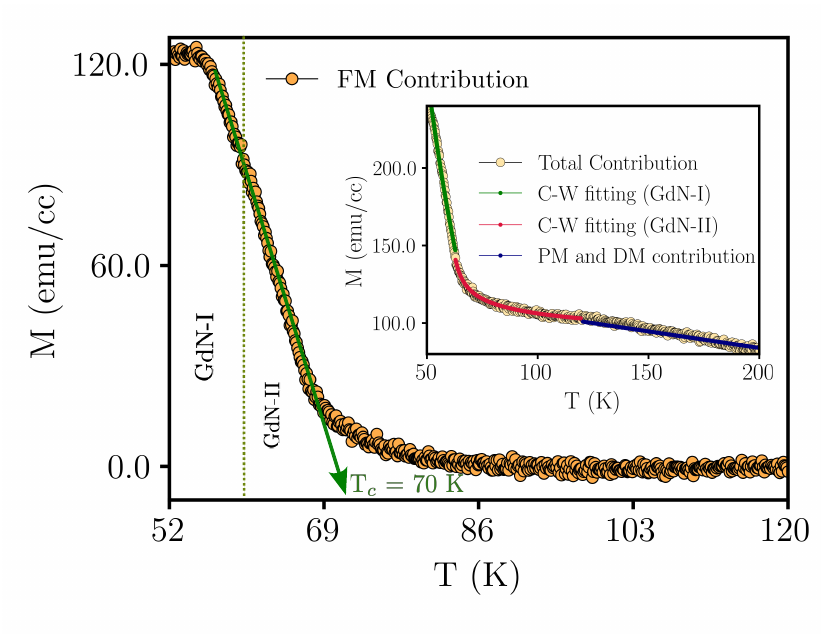} 
\caption{Temperature-dependent magnetization (MT) of the grown GdN thin films, measured under an applied magnetic field of 200 mT, exhibits a Curie temperature (\textit{T$_c$}) of approximately 70 K. Inset depicts the total magnetization contribution, comprising of various magnetic contributions including paramagnetic (PM) and ferromagnetic (FM) phases, fitted by the combination of Curie $\left(\frac{C}{T} \right)$ and Curie-Weiss equation $\left(\frac{C}{T - \theta_P}\right)$, resulting in two GdN phases, I and II. These phases originated due to nitrogen vacancies that have similar lattice parameters and distinctive magnetic properties.}
\label{Raman DFT}
\end{figure}

\section{Magnetic Characteristics}
The magnetic properties of the GdN thin film grown on an AlN/SiO$_2$ substrate were investigated using vibrating sample magnetometry (VSM) through field- and temperature-dependent measurements. Temperature-dependent magnetization ($M$--$T$) measurements were performed in zero-field-cooled (ZFC), field-cooled cooling (FCC), and field-cooled warming (FCW) modes over the temperature range of 48--300~K under an applied magnetic field of 200~mT, as shown in Fig. S4. Figure~4 shows the ferromagnetic (FM) contribution obtained by subtracting the linear contributions consisting of PM contribution from the total FCW magnetization curve, revealing a clear intercept at $\sim70$~K, corresponding to the Curie temperature ($T_\mathrm{c}$) of the GdN thin film, in good agreement with reported values~\cite{Li1994-jx}. An additional anomaly is observed near 124~K, indicating the coexistence of ferromagnetic and paramagnetic (PM) phases, consistent with the presence of a condensed GdN-II phase identified from structural analysis. The magnetization vanishes above $\sim86$~K, effectively excluding metallic Gd impurities, which possess a much higher $T_\mathrm{c}$ of 290~K~\cite{Senapati2010-ax}. The presence of secondary oxide phases such as Gd$_2$O$_3$ is also ruled out, as Gd$_2$O$_3$ exhibits antiferromagnetic behavior~\cite{Mutelet2011-hx}, whereas the observed $M$--$T$ response is ferromagnetic. Furthermore, oxygen-induced magnetization effects are excluded, since increased oxygen incorporation is known to significantly enhance both magnetization and coercivity in GdN thin films~\cite{Senapati2010-ax}. The non-magnetic nature of the AlN buffer layer~\cite{Shi2014-au} confirms that the observed magnetic response originates solely from the GdN film.

The inset of Fig.~4. displays the total FCW magnetization, comprising diamagnetic (DM), PM, and FM contributions. The diamagnetic background arises from the AlN/SiO$_2$ substrate, while the magnetic contributions originate from interacting and non-interacting Gd$^{3+}$ ions within the film. The temperature-dependent magnetization was analyzed using a modified Curie--Weiss formalism,
\begin{equation}
\frac{C}{T - \theta} + \frac{P}{T} + \chi,
\end{equation}
where the first term represents the FM contribution described by the Curie--Weiss model, the second term corresponds to the PM response, and the third term accounts for the diamagnetic background. The low-temperature region (48--63~K) exhibits enhanced $T_\mathrm{c}$ and magnetic moment, attributed to lattice expansion induced by high deposition pressure and reduced N$_2$ flow, resulting in a Gd-deficient, N-rich GdN-I phase~\cite{Sagar2013-sj}. The intermediate region (64--124~K) is characterized by reduced magnetic moment but enhanced $T_\mathrm{c}$, arising from nitrogen-vacancy (\textit{V$_\mathrm{N}$}) formation and the emergence of the condensed GdN-II phase. These vacancies mediate ferromagnetic and antiferromagnetic superexchange interactions between neighboring Gd ions, leading to bound magnetic polaron (BMP) formation and a higher effective $T_\mathrm{c}$~\cite{Natali2013-ok,Shaib2020-ox}. Above 125~K, the magnetic response is dominated by PM and DM contributions with negligible net magnetization.

\begin{figure}[!ht]
\centering
\includegraphics[width=1\linewidth]{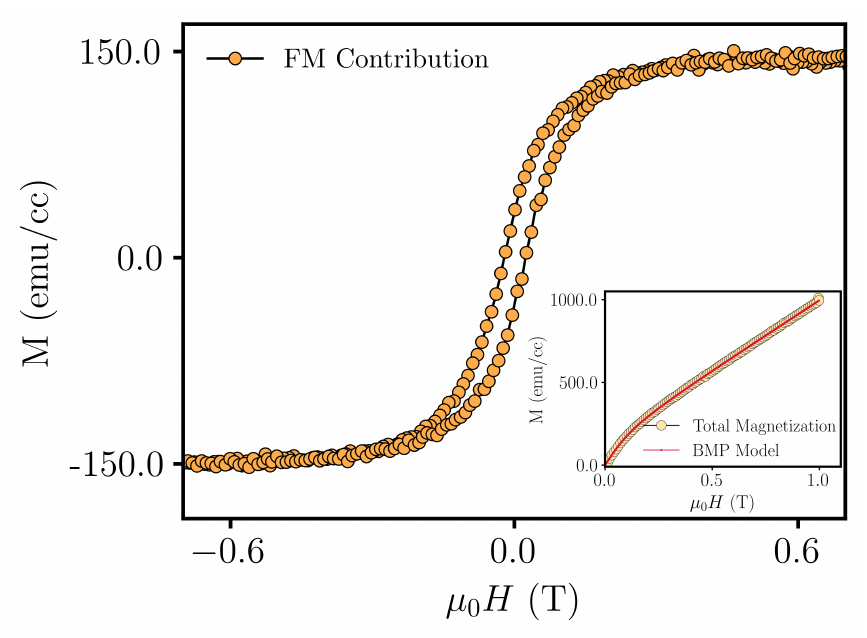} 
\caption{Field-dependent magnetization (MH) of grown GdN thin films at 52 K, depicting a coercivity of 200 Oe and a magnetic moment of 151 emu/cc, confirms the soft magnetic characteristics of GdN. Inset illustrates the virgin MH curve with the influence of nitrogen vacancies with Gd$^{3+}$ ion interaction on the magnetic characteristics of GdN by forming bound magnetic polarons (BMP), resulting in high magnetizations value due to Paramagnetic (PM) contribution. The BMP function was fitted to forward magnetic hysteresis (MH) sweeps using the equation $M = M_0 L(x) + \chi_M H$, where the first term contributes to the effect of BMP density, and the second term represents the PM contribution. }
\label{MH}
\end{figure}

\begin{figure*}[!ht]
\centering
\includegraphics[width=\textwidth]{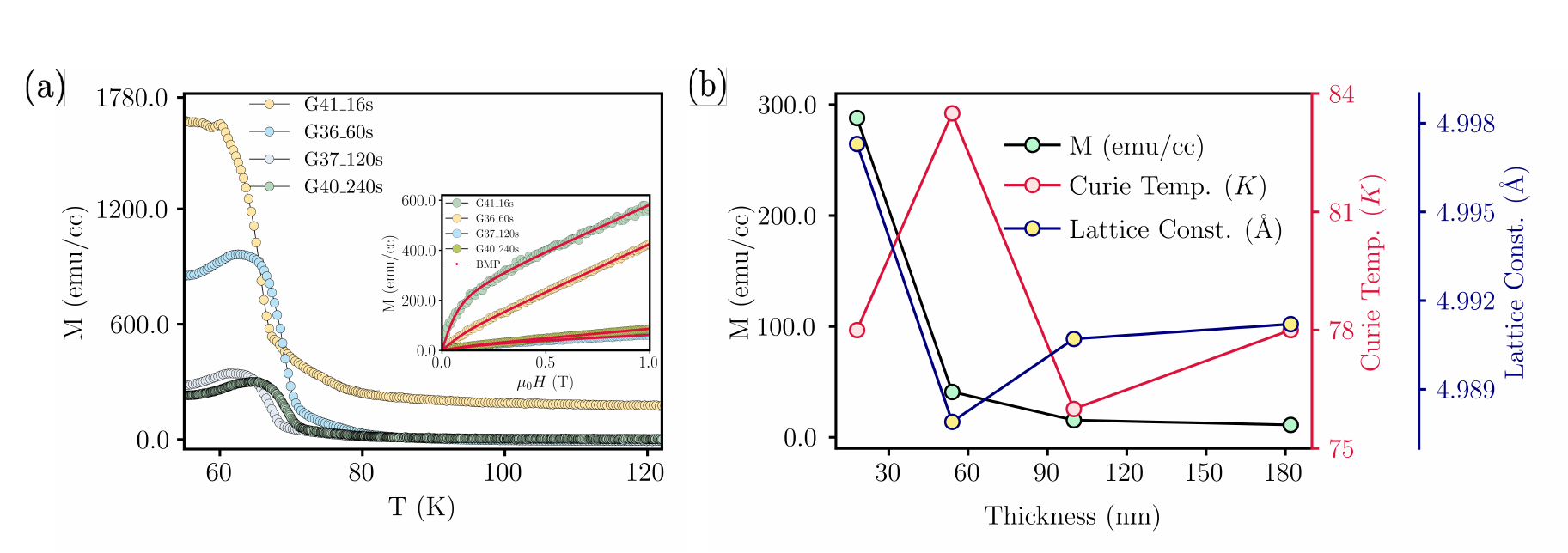} 
\caption{(a) MT graph illustrates various grown GdN samples with varied range of thicknesses spanning from 18 to 182 nm shows the variation in the (\textit{T$_c$}), while inset shows BMP fitting of all MH curves reveals the dependence of  BMP density on reduced magnetization, which directly influenced by nitrogen vacancies (b) illustrates the correlation between thickness, lattice constant, and Curie temperature, as the lattice structure undergoes distortion because of the lattice misfit and strain,  an increase in nitrogen vacancies occurs, subsequently, there is a reduction in magnetization and elevation in (\textit{T$_c$}). }
\label{thickness}
\end{figure*}

Magnetic field–dependent magnetization ($M$–$H$) measurements were carried out using forward and reverse magnetic field sweeps up to $\pm1$~T. Fig.~5 shows the ferromagnetic (FM) contribution extracted from the paramagnetic (PM)-dominated total magnetization at 52~K. The hysteresis loop exhibits a magnetization of 151.2 emu/cc and a coercive field of $\sim220$~Oe, confirming the soft ferromagnetic nature of the grown GdN thin film. The relatively low coercivity further indicates the negligible influence of oxidation, as even small oxygen incorporation is known to increase coercivity to several thousand oersted in GdN thin films~\cite{Holmes-Hewett2025-sw}. In addition, the soft hysteresis behavior suggests a limited contribution from the condensed GdN-II phase, which is otherwise reported to enhance coercivity~\cite{Senapati2011-bq}.

Despite the clear FM response, the saturation magnetization is lower than that of bulk GdN. Possible origins include oxidation, insufficient magnetic field strength, interfacial effects, and nitrogen vacancies (\textit{V$_\mathrm{N}$}) arising from structural disorder. Oxidation effects are excluded based on structural, Raman, and $M$–$T$ analyses, while measurements performed up to $\pm3$~T rule out field-limited magnetization. Prior studies have shown that deviations of lattice parameters from bulk values and associated structural disorder significantly suppress the magnetic moment in GdN thin films~\cite{Khazen2006-ff,Shimokawa2015-xm,Punya2011-gb}. In the present case, high-pressure growth combined with low N$_2$ flow induces lattice strain and disorder, promoting \textit{V$_\mathrm{N}$} formation and restricting the participation of Gd$^{3+}$ ions in long-range ferromagnetic ordering, consistent with earlier XRD, Raman, and magnetization results,  resulting in the lower magnetic moment of $0.22 ~\mu_{\mathrm{B}}$ per Gd$^{3+}$ ion lower then the bulk GdN value. Furthermore, first-principles calculations of pristine GdN with on-site Coulomb interaction parameter $U = 8$~eV yield a local magnetic moment of $7~\mu_{\mathrm{B}}$ per Gd$^{3+}$ ion, which arises from the pronounced spin splitting of electronic bands, as shown in Fig.S5 of SI. In contrast, simulations incorporating a nitrogen vacancy concentration of 3.125\% within a $2 \times 2 \times 2$ supercell reveal a reduced effective global magnetization of $4.3~\mu_{\mathrm{B}}$ per Gd$^{3+}$ ion in GdN lattice, which can be attributed to the stabilization of antiferromagnetic ordering in the \textit{V$_\mathrm{N}$}-rich regions. This theoretical result is consistent with the experimentally observed reduction in magnetization with the effect of \textit{V$_\mathrm{N}$}.

To elucidate the role of \textit{V$_\mathrm{N}$} in governing the magnetic response, the virgin forward-sweep $M$–$H$ curve was analyzed using the bound magnetic polaron (BMP) model, as shown in the inset of Fig.~5. In this framework, \textit{V$_\mathrm{N}$} act as localized carriers that exchange-couple with nearby Gd$^{3+}$ ions, forming FM or AFM polarons embedded in a PM matrix. The magnetization is described by
\begin{equation}
M = M_0 L(x) + \chi_M H,
\end{equation}
where $L(x)=\coth(x)-1/x$ is the Langevin function with $x=\mu_{\mathrm{eff}}H/k_\mathrm{B}T$. The first term represents the FM contribution arising from BMP clusters, while the second term accounts for the PM background. Here, $M_0$ denotes the saturation magnetization, $\mu_{\mathrm{eff}}$ is the effective magnetic moment per BMP cluster, and the BMP density depends sensitively on the concentration of \textit{V$_\mathrm{N}$}. An increased density and overlap of BMPs enhances the Curie temperature but simultaneously reduces the net magnetization due to competing FM and AFM superexchange interactions~\cite{Lee2015-ea,Duan2005-tr}. Accordingly, the disordered GdN-II phase is attributed to \textit{V$_\mathrm{N}$}-mediated BMP formation, which elevates $T_\mathrm{c}$ while suppressing the magnetic moment, in agreement with our $M$–$T$ and $M$–$H$ results and earlier reports~\cite{Wachter2016-rq,Senapati2011-bq}.

\begin{table}[!ht]
\centering
\scriptsize
\caption{Effect of deposition time on structural and magnetic properties}
\label{tab:bmp_params}
\begin{tabular}{|c|c|c|c|c|c|}
\hline
\textbf{Sample ID} &
\makecell{\textbf{Deposition}\\ \textbf{Time (s)}} &
\makecell{\textbf{Dislocation}\\ \textbf{Density}, $\rho$\\ (cm$^{-2}$)} &
\makecell{\textbf{BMP}\\ \textbf{Density}, $N$\\ (cm$^{-3}$)} &
\makecell{\textbf{Effective}\\ \textbf{Moment}, $\mu_{\mathrm{eff}}$\\ (emu)} &
\makecell{\textbf{Molar}\\ \textbf{Susceptibility}, $\chi_m$}
\\ \hline

G41 & 16  & $2.27 \times 10^{9}$ & $9.01 \times 10^{25}$ & $2.49 \times 10^{-24}$ & 0.036 \\ \hline
G36 & 60  & $7.54 \times 10^{9}$ & $2.63 \times 10^{25}$ & $1.77 \times 10^{-24}$ & 0.0038 \\ \hline
G37 & 120 & $1.08 \times 10^{8}$ & $2.29 \times 10^{25}$ & $9.68 \times 10^{-25}$ & 0.0041 \\ \hline
G40 & 240 & $2.13 \times 10^{8}$ & $9.61 \times 10^{24}$ & $1.41 \times 10^{-24}$ & 0.0073 \\ \hline
\end{tabular}
\end{table}

To further elucidate the role of lattice strain and structural disorder on nitrogen vacancy (\textit{V$_\mathrm{N}$}) formation, a series of GdN thin films were deposited by keeping all growth parameters constant while varying only the deposition time from 15 to 240~s, yielding film thicknesses in the range of 18--183~nm. All samples exhibit preferential (111) orientation, as confirmed by the $2\theta$--$\omega$ scans shown in Fig.~S4(a) of the Supplementary Information. With increasing deposition time, an additional lattice mismatch of $\sim2.87\%$ develops between the substrate and the GdN film, introducing enhanced lattice strain and structural disorder. Table~I demonstrates a strong correlation between deposition time, film thickness, and threading dislocation density. The dislocation density initially increases with thickness up to sample G36 and subsequently decreases for thicker films, indicating that thinner GdN layers possess a higher density of defects and disorder, which promotes increased \textit{V$_\mathrm{N}$} formation up to a critical thickness. Beyond this threshold, partial N$_2$ enrichment within interstitial sites leads to defect-ridden crystalline growth with reduced dislocation density. Fig.~6~(a) shows the field-cooled cooling (FCC) magnetization--temperature ($M$--$T$) curves for GdN films grown with different deposition times, highlighting systematic variations in magnetization and Curie temperature ($T_\mathrm{c}$). With increasing thickness, the overall magnetization decreases, while the irreversibility temperature associated with the GdN-II phase---identified by the bifurcation of ZFC and FCC curves (Fig.~S~4(b)---exhibits a clear dependence on deposition time. Fig.~6~(b) correlates $T_\mathrm{c}$ with film thickness and lattice constant, demonstrating that increased lattice misfit and strain enhance the dislocation density, which scales with the Burgers vector~\cite{Plank2011-po}. Consequently, sample G40, which exhibits the highest dislocation density, also shows the maximum $T_\mathrm{c}$. The inset of Fig.~6~(a) presents bound magnetic polaron (BMP) fits to the $M$--$H$ data for all films, revealing a systematic reduction in the effective magnetic moment per BMP with increasing thickness, as summarized in Fig.~6~(b). The reduction in magnetization is attributed to a decrease in BMP density, reflecting a lower concentration of \textit{V$_\mathrm{N}$} interacting with Gd$^{3+}$ ions and, hence, smaller BMP cluster sizes. Nevertheless, the magnetization remains significantly lower than the bulk GdN value due to the coexistence of ferromagnetic and antiferromagnetic exchange interactions mediated by \textit{V$_\mathrm{N}$}, which suppress the net magnetic moment. Importantly, the reduced magnetization and low coercivity observed in these thin films are advantageous for rapid magnetic switching, highlighting their potential suitability for spintronic device applications.
To further elucidate the role of lattice strain and structural disorder on nitrogen vacancy (\textit{V$_\mathrm{N}$}) formation, a series of GdN thin films were deposited by keeping all growth parameters constant while varying only the deposition time from 15 to 240~s, yielding film thicknesses in the range of 18--183~nm. All samples exhibit preferential (111) orientation, as confirmed by the $2\theta$--$\omega$ scans shown in Fig.~S4~(a) of the SI. With increasing deposition time, an additional lattice mismatch of $\sim2.87~\%$ develops between the substrate and the GdN film, introducing enhanced lattice strain and structural disorder. Table~I demonstrates a strong correlation between deposition time, film thickness, and threading dislocation density. The dislocation density initially increases with thickness up to sample G36 and subsequently decreases for thicker films, indicating that thinner GdN layers possess a higher density of defects and disorder, which promotes increased \textit{V$_\mathrm{N}$} formation up to a critical thickness. Beyond this threshold, partial N$_2$ enrichment within interstitial sites leads to defect-ridden crystalline growth with reduced dislocation density. Fig.~6~(a) shows the field-cooled cooling (FCC) magnetization--temperature ($M$--$T$) curves for GdN films grown with different deposition times, highlighting systematic variations in magnetization and Curie temperature ($T_\mathrm{c}$). With increasing thickness, the overall magnetization decreases, while the irreversibility temperature associated with the GdN-II phase---identified by the bifurcation of ZFC and FCC curves (Fig.~S4~(b))---exhibits a clear dependence on deposition time. Fig.~6~(b) correlates $T_\mathrm{c}$ with film thickness and lattice constant, demonstrating that increased lattice misfit and strain enhance the dislocation density, which scales with the Burgers vector~\cite{Plank2011-po}. Consequently, sample G40, which exhibits the highest dislocation density, also shows the maximum $T_\mathrm{c}$. The inset of Fig.~6~(a) presents bound magnetic polaron (BMP) fits to the $M$--$H$ data for all films, revealing a systematic reduction in the effective magnetic moment per BMP with increasing thickness, as summarized in Fig.~6~(b). The reduction in magnetization is attributed to a decrease in BMP density, reflecting a lower concentration of \textit{V$_\mathrm{N}$} interacting with Gd$^{3+}$ ions and, hence, smaller BMP cluster sizes. Nevertheless, the magnetization remains significantly lower than the bulk GdN value due to the coexistence of ferromagnetic and antiferromagnetic exchange interactions mediated by \textit{V$_\mathrm{N}$}, which suppress the net magnetic moment. Importantly, the reduced magnetization and low coercivity observed in these thin films are advantageous for rapid magnetic switching, highlighting their potential suitability for spintronic device applications.

\section{Conclusions}

 A comprehensive structural, spectroscopic, and magnetic characterization of sputtered GdN thin films grown on AlN/SiO$_2$ substrates reveals polycrystalline cubic GdN with preferential (111) orientation and controlled lattice strain. XRD, AFM, XPS, and Raman analyses establish high crystalline quality, moderate surface roughness, and the presence of nitrogen-vacancy–induced symmetry breaking without detectable oxide phases. Temperature- and field-dependent magnetometry confirms ferromagnetic ordering with a Curie temperature near 70 K and an enhanced magnetic phase up to $\sim$125 K associated with a condensed, vacancy-rich GdN-II phase. Systematic thickness variation demonstrates that lattice strain and dislocation density govern nitrogen-vacancy formation, which mediates bound magnetic polaron interactions, leading to enhanced $T_\mathrm{c}$ but reduced magnetization.Furthermore, the significant correlation between the observed experimental outcomes and the simulated theoretical results underscores the influence of the nitrogen vacancy on the dynamics of ferromagnetic GdN. These results establish nitrogen vacancies as the central parameter controlling structure–magnetism correlations in GdN thin films and highlight their relevance for low-coercivity spintronic applications.

  Authors are thankful to the National Nano-Fabrication Facility (NNFC) and the Micro and Nano Characterization Facility (MNCF) at the Centre for Nanoscience and Engineering, IISc.  PB thanks Anusandhan National Research Foundation (ANRF), National Postdoctoral fellowship (PDF/2023/000444) for financial support. JS acknowledges JNCASR for research fellowship. UVW acknowledges suppoort from a J.C. Bose Grant of ANRF, Govt. of India. DS thanks IISc start-up grant, Ministry of Electronics and Technology, Indian Space Research Organization for funding. Authors duly acknowledge funding from INOXCVA and INOX Airproducts for funding via CSR grants.

\bibliographystyle{apsrev4-2}
\bibliography{References}
\end{document}

\def\scititle{Supplementary Text for \\ Nitrogen-Vacancy-Mediated Magnetism in Sputtered GdN Thin Films}
\title{\bfseries \boldmath \scititle}

\author{Pankaj Bhardwaj}
\email{pankajb@iisc.ac.in}
\affiliation{Centre for Nanoscience and Engineering, Indian Institute of Science, Bengaluru 560012, India} 
\author{Jyotirmoy Sarkar}
\affiliation{Theoretical Sciences Unit, Jawaharlal Nehru Center for Advanced Scientific Research, Jakkur, Bengaluru 560064, India}
\author{Bubun Biswal}
\affiliation{Department of Physics, Indian Institute of Technology Madras, Chennai 600036, India}
\author{Subhransu Kumar Negi}
\affiliation{Centre for Nanoscience and Engineering, Indian Institute of Science, Bengaluru 560012, India}
\author{Arijit Sinha}
\affiliation{Theoretical Sciences Unit, Jawaharlal Nehru Center for Advanced Scientific Research, Jakkur, Bengaluru 560064, India}
\author{Anirudh Venugopalrao }
\affiliation{Centre for Nanoscience and Engineering, Indian Institute of Science, Bengaluru 560012, India}
\author{Sharath Kumar C}
\affiliation{School of Physics, IISER Thiruvananthapuram, Vithura, Thiruvananthapuram 695551, India}
\author{Sreelakshmi M Nair}
\affiliation{Department of Physics, BITS Pilani K. K. Birla Goa Campus, Zuarinagar 403726, Goa, India}
\author{R. S. Patel}
\affiliation{Department of Physics, BITS Pilani K. K. Birla Goa Campus, Zuarinagar 403726, Goa, India}
\author{Deepshika Jaiswal Nagar}
\affiliation{School of Physics, IISER Thiruvananthapuram, Vithura, Thiruvananthapuram 695551, India}
\author{Abhishek Mishra}
\affiliation{Department of Physics, Indian Institute of Technology Madras, Chennai 600036, India}
\author{Srinivasan Raghavan}
\affiliation{Centre for Nanoscience and Engineering, Indian Institute of Science, Bengaluru 560012, India}
\author{Umesh Waghmare}
\affiliation{Theoretical Sciences Unit, Jawaharlal Nehru Center for Advanced Scientific Research, Jakkur, Bengaluru 560064, India}
\author{Dhavala Suri}
\email{dsuri@iisc.ac.in}
\affiliation{Centre for Nanoscience and Engineering, Indian Institute of Science, Bengaluru 560012, India}

\maketitle
\newpage
\renewcommand{\thesection}{S\arabic{section}}
\setcounter{section}{0}

\section{\NoCaseChange{Deposition growth parameters}}

\noindent The growth of thin films in a sputtering system is influenced by several key factors, including base vacuum pressure, deposition pressure, substrate temperature, and the distance between the substrate and the target. Additionally, the Argon (Ar) and Nitrogen (N$_2$) flow rates, their respective ratios, and the type of power supply, deposition power, time, and pre- and post-deposition parameters all play significant roles in determining the film growth rate and characteristics. Hence, controlling and optimizing all the parameters make the growth of GdN thin films more complicated. Another challenge in the growth of GdN thin films is its high oxophilicity. Gadolinium (Gd) readily reacts with oxygen to form gadolinium oxide (Gd$_2$O$_3$), requires meticulous control over synthesis. The Table S1 shows the deposition parameters of all grown GdN samples experimented in this word

% Insert your text or table for growth parameters here
\begin{table}[!ht]
\centering
\scriptsize
{\normalsize\noindent Table S1. Deposition parameters of all samples experimented on in this work.}

%\label{tab:deposition_params}

\resizebox{\textwidth}{!}{
\begin{tabular}{|c|c|c|c|c|c|c|c|c|c|c|}
\hline
\textbf{Sample ID} & \textbf{Substrate} & \makecell{\textbf{Substrate}\\ \textbf{Temp}\\(K)} & 
\makecell{\textbf{Base}\\ \textbf{Pressure}\\(mbar)} & 
\makecell{\textbf{Dep.}\\ \textbf{Time}\\(s)} &
\makecell{\textbf{DC}\\ \textbf{Power}\\(W)} &
\makecell{\textbf{Ar and N$_2$}\\ \textbf{Flow}\\(sccm)} &
\makecell{\textbf{Target–Substrate}\\ \textbf{Distance}\\(cm)} &
\makecell{\textbf{Post Annealing}\\ \textbf{Gas (N$_2$)}\\ (sccm)} &
\makecell{\textbf{Nitridation}\\ \textbf{Temp}\\($^\circ$C)} &
\makecell{\textbf{Nitridation}\\ \textbf{Duration}\\(hr)} \\
\hline 
G24 & AlN/SiO$_2$ & 500 & $3.31 \times 10^{-7}$ & 60 & 100 & 7N$_2$ - 5Ar & $\approx 7.8$ & - & - & - \\
\hline
G43 & AlN/SiO$_2$ & 500 & $3.05 \times 10^{-7}$ & 60 & 100 & 7N$_2$ - 4Ar & $\approx 7.8$ & 50 & $500^\circ$ & 1 \\
\hline
G36 & AlN/SiO$_2$ & 500 & $2.89 \times 10^{-7}$ & 60 & 100 & 7N$_2$ - 4Ar & $\approx 7.8$ & 50 & $500^\circ$ & 1 \\
\hline
G37 & AlN/SiO$_2$ & 500 & $2.68 \times 10^{-7}$ & 120 & 100 & 7N$_2$ - 4Ar & $\approx 7.8$ & 50 & $500^\circ$ & 1 \\
\hline
G40 & AlN/SiO$_2$ & 500 & $2.99 \times 10^{-7}$ & 240 & 100 & 7N$_2$ - 4Ar & $\approx 7.8$ & 50 & $500^\circ$ & 1 \\
\hline
G41 & AlN/SiO$_2$& 500 & $3.01 \times 10^{-7}$ & 18 & 100 & 7N$_2$ - 4Ar & $\approx 7.8$ & 50 & $500^\circ$ & 1 \\
\hline
\end{tabular}
}
\end{table}

The G24 and G43 sample codes are used to compare the growth of GdN, as shown in Fig. 1, while G36, G37, G40, and G41 are used for thickness-dependent studies.
\clearpage

\renewcommand{\thesection}{S\arabic{section}}
\setcounter{section}{1}

\section{\NoCaseChange{The Omega Scan}}

% Insert your Omega scan figure here using \begin{figure} ... \end{figure}
\begin{figure*}[!ht]
    \centering
    \includegraphics[width=0.8\textwidth]{SI1.pdf} % Replace with your filename
    
    \noindent \justifying Fig.~S1:~ The $\omega$-scan of the GdN (111) reflection, fitted with a Gaussian function having a full width at half maximum (FWHM) of 0.12, confirmed the crystalline, textured thin films with an incursion of dislocation density of 2.84 $\times$ 10\textsuperscript{9} cm\textsuperscript{2}
    \label{intro}
\end{figure*}

\clearpage

\renewcommand{\thesection}{S\arabic{section}}
\setcounter{section}{2}

\section{\NoCaseChange{XPS survey scan and High-resolution Gd and N  atom XPS scan}}

% Insert your XPS figures here
\begin{figure*}[!ht]
    \centering
    \includegraphics[width=\textwidth]{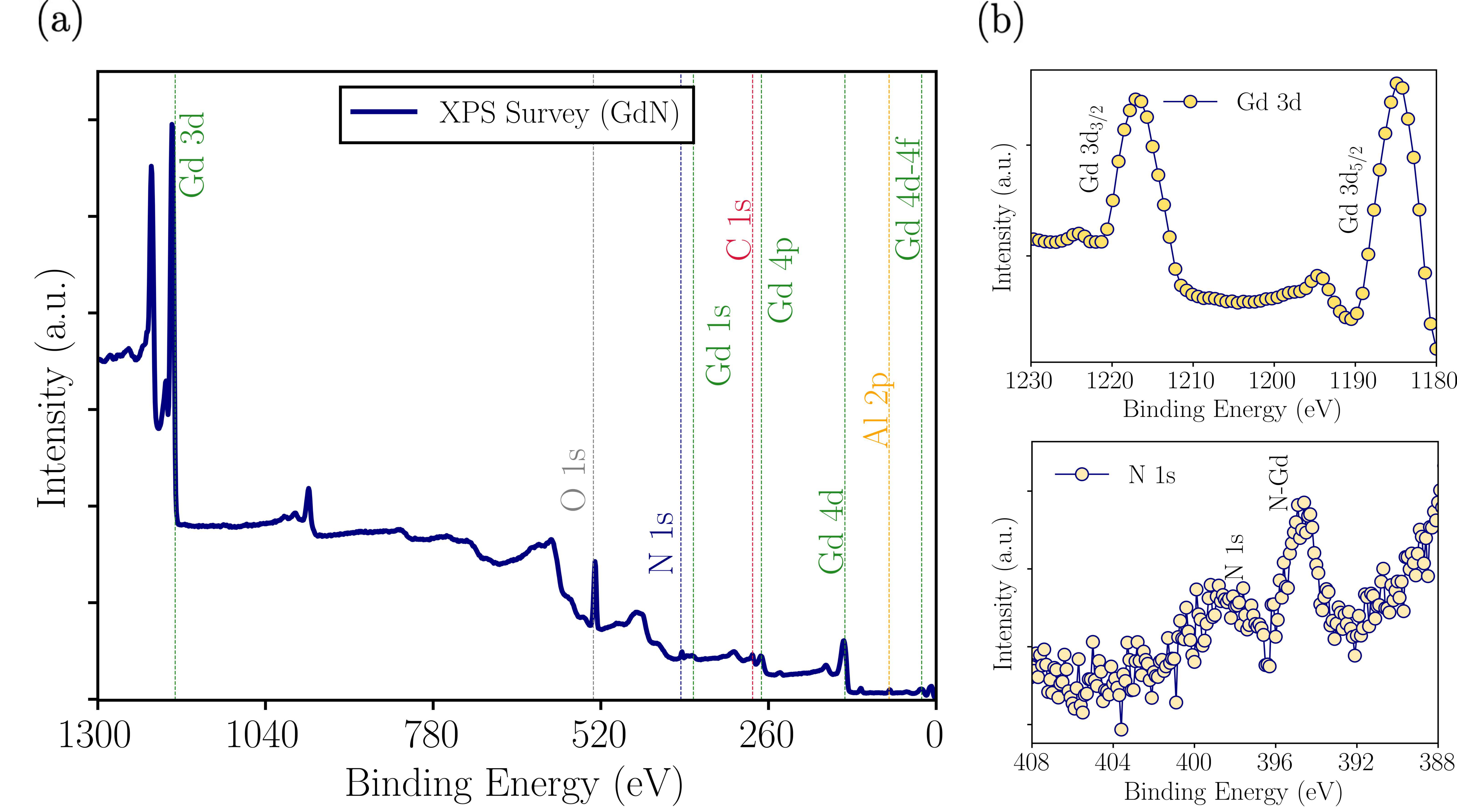}
\noindent\justifying Fig.~S2:~ (a) The XPS survey of grown GdN thin film from 0 to 1300 eV, and (b) high-resolution XPS Gd 3d and N 1s  scan.
       \label{intro}
\end{figure*}
 The XPS survey shows the Gd 3d, N 1s, Al 2p, C 1s and O 1s element in the grown GdN thin films. Residual carbon and oxygen signals are also observed in the etched spectra, which are attributed to unavoidable atmospheric exposure during sample transfer. This behavior is consistent with the known oxophilic nature of GdN and the absence of surface passivation, which leads to rapid oxidation upon air exposure. To mitigate this effect, all GdN films were stored in a high-pressure N$_2$ environment prior to characterization. Importantly, XRD measurements show no evidence of secondary oxide phases, which confirms the phase purity of the films. The high resolution XPS spectra reveals the splitting of Gd 3d element into 3d$_{3/2}$ and 3d$_{5/2}$ positions at 1220 and 1185 eV, respectively. In contrast, the N-1s core level spectra observed at 395 eV exhibit a shift of 2 eV from 398 eV, attributed to the electronegativity difference between Gd and N atoms, which confirm the Gd-N interaction.

\clearpage

\renewcommand{\thesection}{S\arabic{section}}
\setcounter{section}{3}

\section{\NoCaseChange{Temperature dependent Raman Measurements with and without applied magnetic field}}

% Insert your Raman figures here
\begin{figure*}[!ht]
    \centering
    \includegraphics[width=\textwidth]{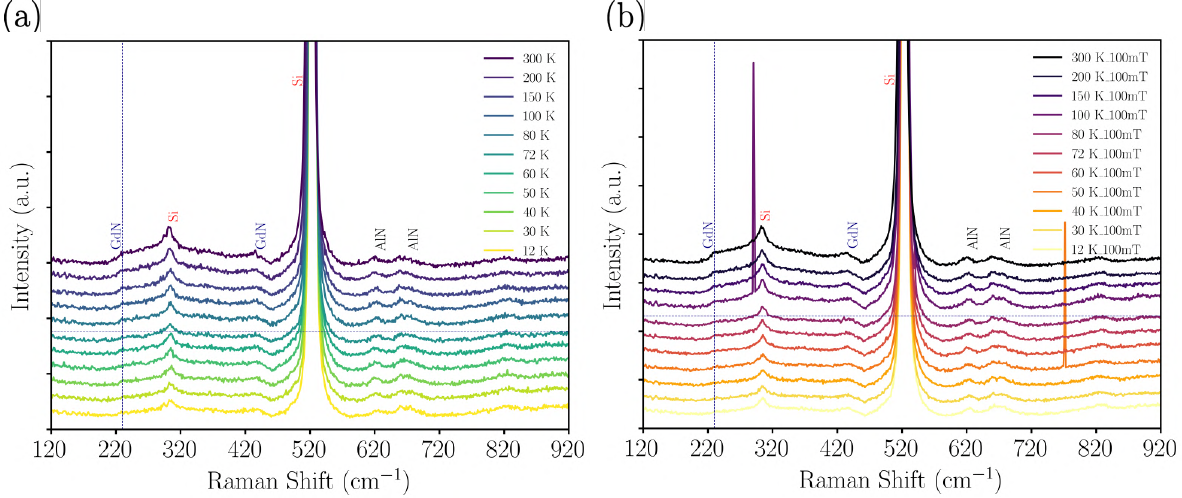} % Replace with your filename
    \noindent \justifying Fig.~S3:~ Temperature dependent Raman spectral measurement of grown GdN thin film (a) without and (b) with applied magnetic field of 100 mT
    \label{intro}
\end{figure*}

The temperature-dependent Raman spectrum measurement demonstrates the flattening of the GdN Raman mode observed at approximately 230 cm$^{-1}$. This flattening is accompanied by a decrease in temperature from 300 K to 12 K, both in the presence and absence of a magnetic field of 100 mT. The Figure unequivocally indicates that 72 K serves as the critical temperature for the flattening of the Raman mode, thereby confirming the Curie temperature of GdN at 72 K.

\clearpage

\renewcommand{\thesection}{S\arabic{section}}
\setcounter{section}{4}

\section{\NoCaseChange{Structural and Magnetization measurement with varying deposition time}}

% Insert your XRD figures here
\begin{figure*}[!ht]
    \centering
    \includegraphics[width=\textwidth]{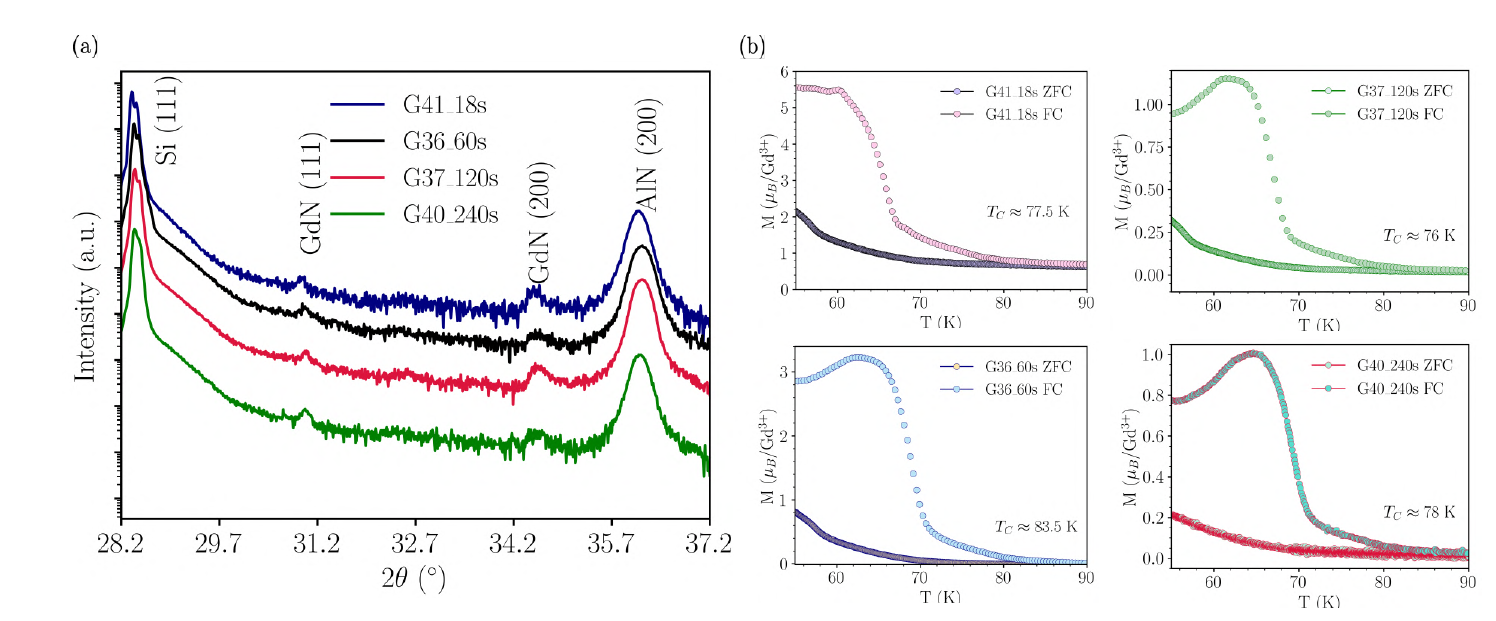} % Replace with your filename
    \noindent \justifying Fig.~S4:~(a) XRD curves (b) MT curve consisting of FC and ZFC curve of GdN grown thin film with varying deposition time.
    \label{intro}
\end{figure*}
Fig.~S4~(a) presents the $2\theta$--$\omega$ XRD scans, which corroborate the preferential growth of GdN in all the deposited thin films. A systematic shift of the diffraction peaks toward lower $2\theta$ values is observed as the film thickness increases up to a critical thickness, followed by a shift toward higher $2\theta$ values. The GdN film deposited for 18~s exhibits a lattice mismatch of $13.68~\%$, which is $2.87~\%$ higher than the bulk value of $13.2\%$, thereby confirming that reduced thickness induces enhanced strain in the grown GdN thin film. As the deposition time increases from 18 to 60~s, the XRD peak shifts toward higher $2\theta$ values, indicating that additional tensile stress, beyond the lattice mismatch, further reduces the lattice constant and results in increased microstrain in the GdN heterostructure. Subsequently, with an increase in deposition time from 60 to 240~s, the diffraction peak shifts toward lower $2\theta$ values, signifying the induction of compressive strain and the onset of strain relaxation toward the bulk lattice parameter.

\noindent \justifying Fig.~S4~(b) illustrates the zero-field-cooled (ZFC) and field-cooled (FC) bifurcation of the raw magnetization–temperature (MT) curves, revealing a variation in the Curie temperature ($T_{\mathrm{c}}$) ranging from 76 to 84~K. This variation is directly correlated with the induced strain and the presence of nitrogen vacancies in the GdN thin films.

\clearpage

\renewcommand{\thesection}{S\arabic{section}}
\setcounter{section}{5}

\section{\NoCaseChange{First Principles Density Functional Theory calculations with and without the efffect of Nitrogen vacancies}}

% Insert your XRD figures here
\textbf{(a)~~Computational Details}\\[0.25cm]
{\fontsize{12pt}{14pt}\selectfont}
\noindent \justifying First-principles density functional theory (DFT) calculations were performed using the  Vienna Ab initio Simulation Package (VASP) code$^{[1-2]}$. Interaction between electrons and ions was treated using projector-augmented-wave (PAW) potentials. A plane–wave basis set was employed to represent the Kohn–Sham (KS) states with an energy cutoff of 590~eV. The exchange–correlation functional was treated within  and Perdew–Burke–Ernzerhof (PBE) functional$^{[3]}$. Brillouin zone integrations were sampled using a $6 \times 6 \times 6$ $k$-point mesh for bulk calculations and a $3 \times 3 \times 3$ mesh for supercell calculations, with electronic occupation smearing of 0.05~eV. Hubbard $U$ correction$^{[4]}$ was included using the rotationally invariant scheme, with $U = 8$~eV applied to Gd $5d$ orbitals$^{[5]}$. Self-consistent field (SCF) energy convergence threshold was set to $10^{-8}$~eV per cell, and Hellmann–Feynman forces on each atom were converged to less than $10^{-3}$~eV/\AA. Phonon dispersion was obtained using the frozen phonon method with a $2 \times 2 \times 2$ supercell for bulk GdN and Density functional perturbation theory (DFPT) using a $1 \times 1 \times 1$ cell for defected GdN cell. Primitive-cell lattice parameters of GdN were 3.52/3.60~\AA{} (unrelaxed/relaxed).

\begin{figure*}[!ht]
    \centering
    \includegraphics[width=\textwidth]{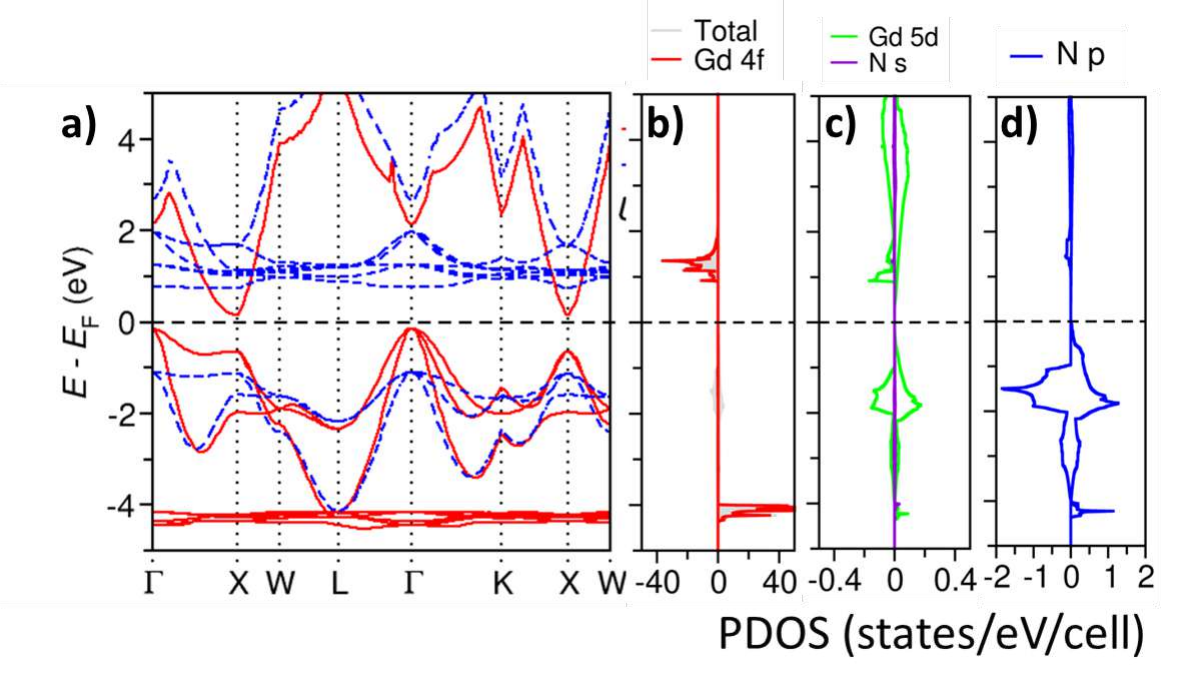} % Replace with your filename
    \noindent Fig.~S5:~Electronic Structure and Projected Density of States (PDOS) of bulk GdN.
    \label{intro}
\end{figure*}

{\fontsize{12pt}{14pt}\selectfont}

\noindent \justifying Benchmarking of the computational methodology was performed first with electronic structure of bulk GdN. Calculated spin-polarized band structure reproduces the expected semiconducting ground state, consistent with earlier reports$^{[6]}$. The valence band dispersion along the high-symmetry directions and bandwidth are in good agreement with earlier reports (In fig. 1, dotted line is Fermi level). However, deviations are observed in the conduction bands, which is attributed to differences in the choice of Hubbard $U$, exchange--correlation functional, and estimates of localized $f$ and $d$ states.
Calculated spin-resolved band gaps are 0.29~eV and 1.85~eV for spin up and down channels respectively. 

\textbf{(b)~~Phonon Mode Visualization}\\[0.25cm]

\begin{figure}[ht!]
    \centering
    \includegraphics[width=0.5\textwidth]{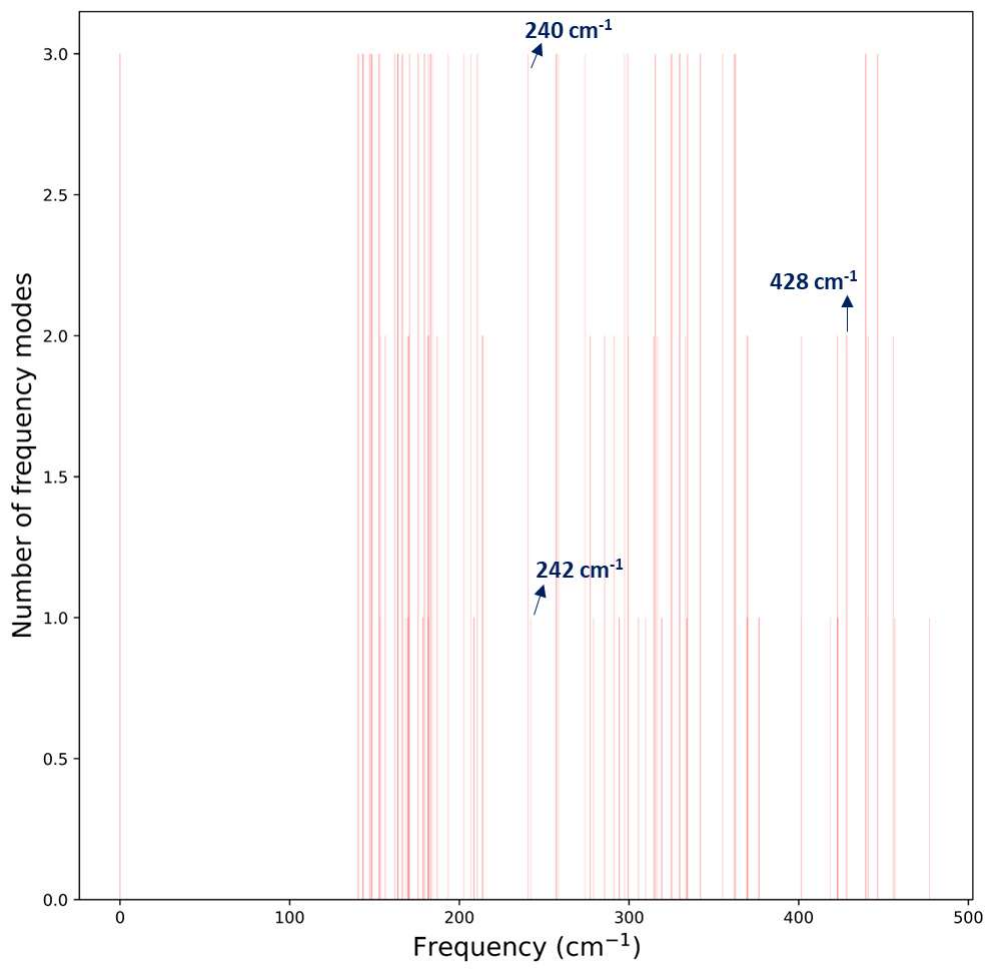}

     \noindent Fig.~S6:~$\Gamma$-point phonon modes of defected GdN
    \label{fig:placeholder}
\end{figure}

\noindent \justifying Here are the visualizations of the 240~cm$^{-1}$, 242~cm$^{-1}$ and 428~cm$^{-1}$ phonon modes of the nitrogen deficient GdN. Visualization arrows represent direction of a specific ion.\\[0.3cm]

\begin{figure}[h!]
    \centering
    \includegraphics[width=\textwidth]{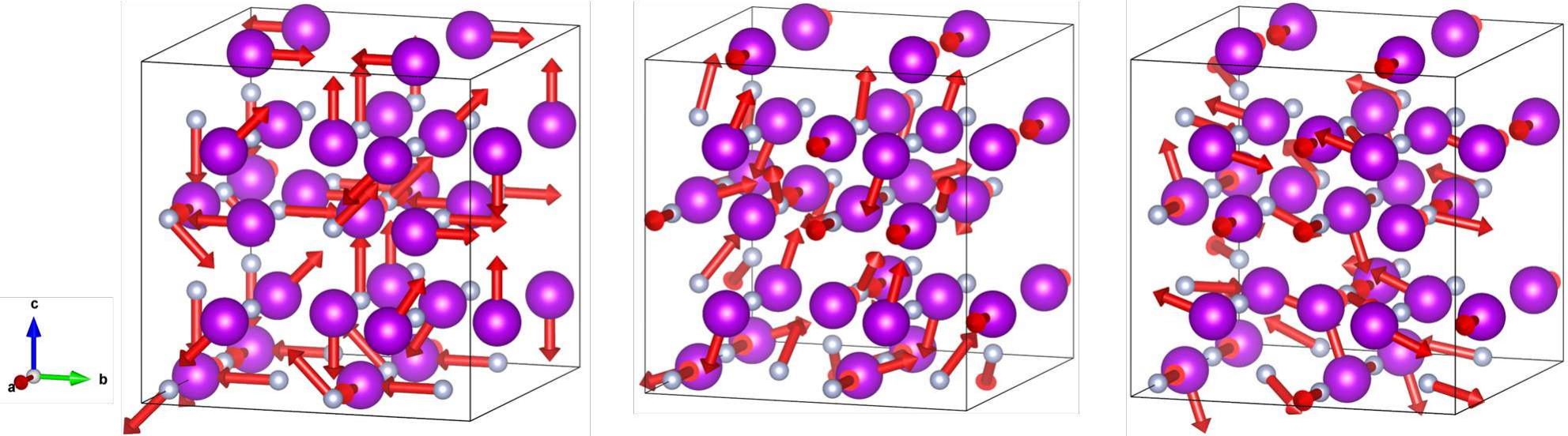}

    \noindent Fig.~S7:~240~cm$^{-1}$ phonon mode
    \label{fig:placeholder}
\end{figure}

\begin{figure}[h!]
    \centering
    \includegraphics[width=0.5\textwidth]{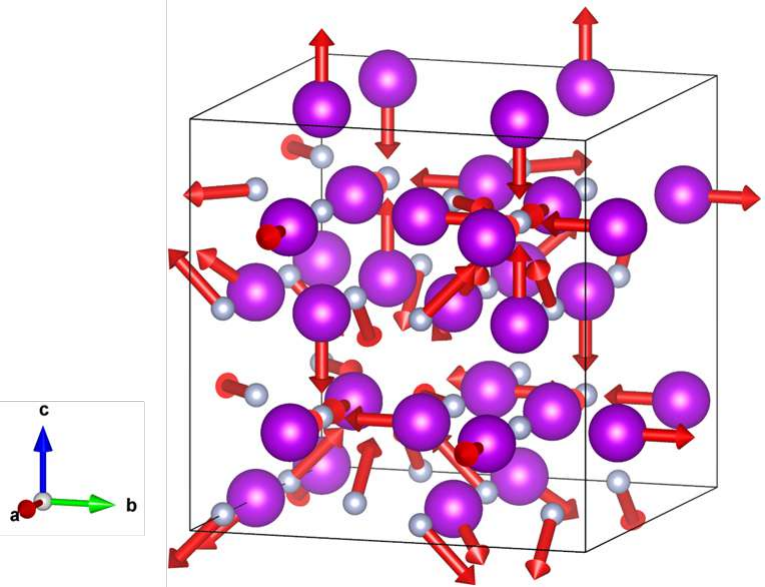}

     \noindent Fig.~S8:~242~cm$^{-1}$ phonon mode
    \label{fig:placeholder}
\end{figure}

\begin{figure}[h!]
    \centering
    \includegraphics[width=\textwidth]{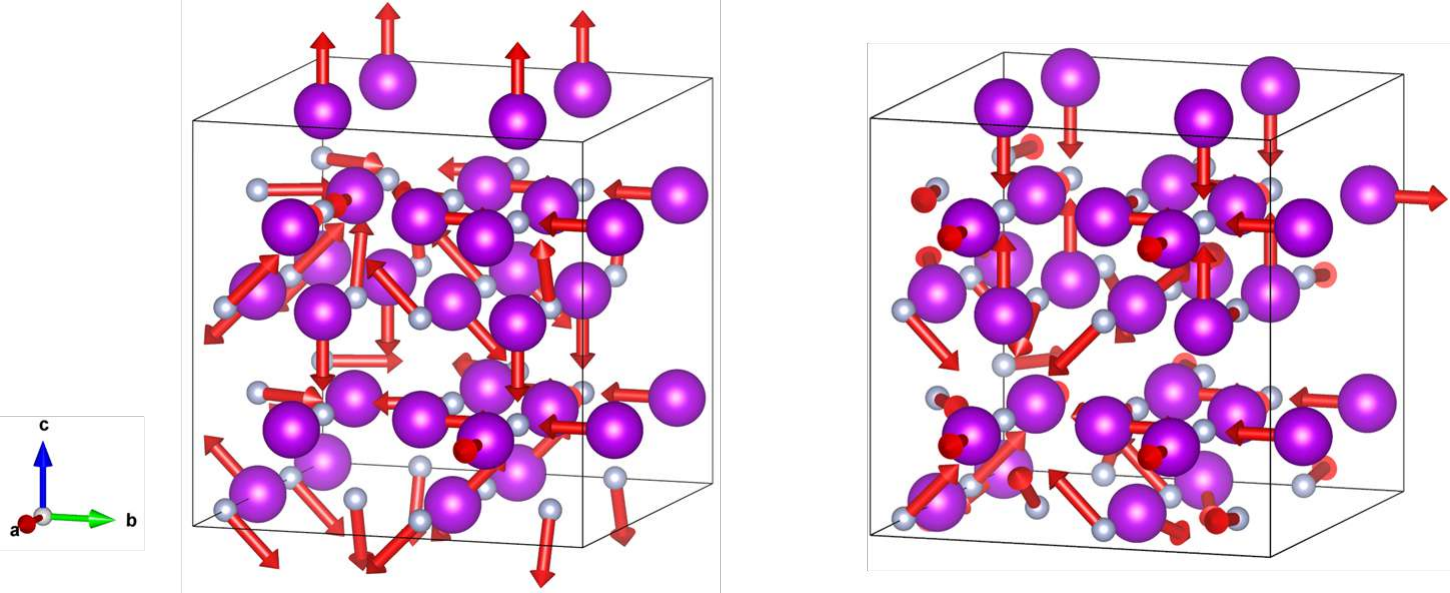}

     \noindent Fig.~S9:~428~cm$^{-1}$ phonon mode
    \label{fig:placeholder}
\end{figure}

\newpage

\noindent \justifying Since defects can absorb or supply crystal momentum, the strict momentum conservation selection rule (requiring near-zero phonon momentum) is relaxed$^{[12]}$. This allows phonons from across the entire Brillouin zone to contribute to the Raman scattering process.

\textbf{(c)~~Magnetic Moment Calculations}\\[0.25cm]
\noindent \justifying First-principles Calculations of Pristine GdN (Space Group: Fm-3m) with $U = 8$~eV give a local magnetic moment of 7~$\mu_{\mathrm{B}}$/Gd$^{3+}$, due to large spin splitting of bands. \\

\begin{figure}[h!]
    \centering
     \includegraphics[width=0.5\textwidth]{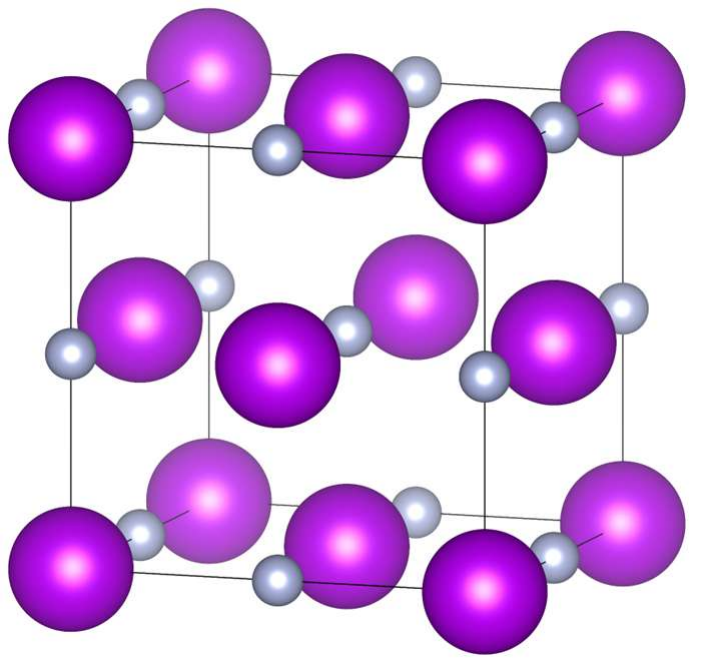}

     \noindent \justifying Fig.~S10:~Conventional unit cell of pristine GdN (Violet colored spheres are Gd atoms and ash colored spheres are N atoms).
    \label{fig:placeholder}
\end{figure}

\noindent \justifying GdN with 3.125\% nitrogen vacancy simulated wih a $2 \times 2 \times 2$ supercell exhibits an effective local magnetic moment of 4.37~$\mu_{\mathrm{B}}$/Gd$^{3+}$.
\begin{figure}[h!]
    \centering
     \includegraphics[width=0.5\textwidth]{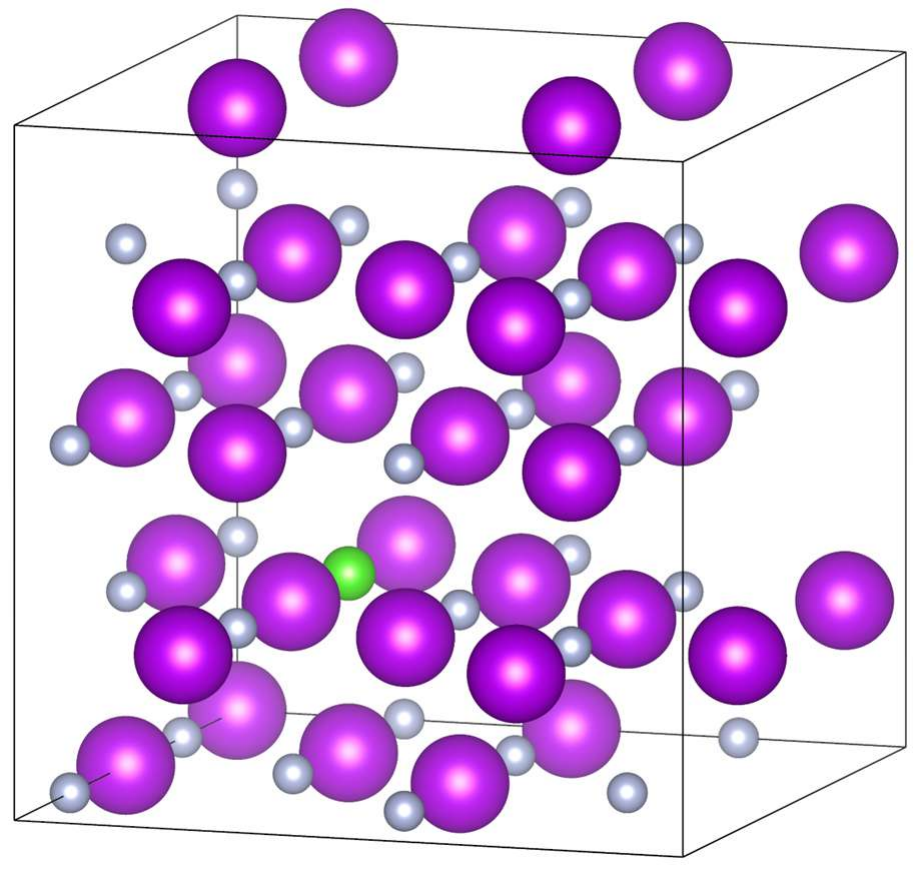}

     \noindent \justifying Fig.~S11:~$2 \times 2 \times 2$ supercell of nitrogen deficient GdN (Green colored sphere is vacancy)
    \label{fig:placeholder}
\end{figure}
\clearpage